\documentclass[11pt]{article}
\usepackage{graphicx}
\begin{document}

\title{Anthropic tuning of the weak scale and of $m_u/m_d$ in two-Higgs-doublet models}

\author{{\bf S.M. Barr and Almas Khan}\\Bartol Research Institute
\\University of Delaware\\Newark, DE 19716}

\date{}
\maketitle

\begin{abstract}

It is shown that in a model in which up-type and down-type fermions acquire mass from
different Higgs doublets, the anthropic tuning of the Higgs mass parameters can explain
the fact that the observed masses of the $d$ and $u$ quarks are nearly the same with
$d$ slightly heavier.  If Yukawa couplings are assumed not to ``scan" (vary among domains),
this would also help explain why $t$ is much heavier than $b$. It is also pointed out that
the existence of dark matter invalidates some earlier anthropic arguments against
the viability of domains where the Standard Model Higgs has positive $\mu^2$, but makes other
even stronger arguments possible.

\end{abstract}

\section{Introduction}

The mass parameter of the Higgs field
in the Standard Model ($\mu^2$) gives the appearance of being ``anthropically tuned" \cite{abds}.
That is, if one imagines the other parameters of the Standard Model to be
fixed, and considers what the universe would look like for different values of $\mu^2$, one finds
that organic life may only be possible if $\mu^2$ is negative and has a magnitude very close
to the value actually observed.  From the physics point of view this might be just a
coincidence, though a remarkable one. On the other hand, as pointed out by Ref. \cite{abds},
it might have a physical explanation
in the context of ``many-domain", ``multiverse", ``landscape" scenarios. (These are equivalent names for the same idea.  For recent reviews see \cite{landscape}.)  An explanation of that sort would account for the closeness of the strong-interaction and
weak-interaction scales ($M_{strong}/M_{P \ell} \sim 10^{-19}$ and $M_{weak}/M_{P \ell} \sim 10^{-17}$), something
not yet explained by any other scenario.  (The nearness of the weak-scale to the strong scale
results in the lightest quarks having masses small compared to $\Lambda_{QCD}$, which in turn leads
to the existence of nonperturbative bound states of quarks, pseudogoldstone pions, and the
the richness of hadronic and nuclear physics, and consequently of chemistry.)

In this paper we extend the analysis of \cite{abds} to a slightly more general Higgs structure and
show that a possible anthropic explanation of the fact that $m_u/m_d \sim 1$ emerges. We also
reconsider the crucial case of positive $\mu^2$, and point out that one of the anthropic arguments proposed
in \cite{abds} to exclude most of the $\mu^2>0$ region is invalidated by the existence of dark energy
(which had not yet been discovered when \cite{abds} was written). However, we show that the existence
of dark energy makes possible a very different set of arguments that reach even stronger conclusions.

The papers of Ref. 1 assumed that the only parameter of the Standard Model that varies among domains
and is anthropically tuned is $\mu^2$. Several facts suggest the possibility that the observed value of $\mu^2$ may be acounted for in a different
way than the values of the other parameters of the Standard Model.
First, $\mu^2$ is the
only dimensionful parameter of the Standard Model Lagrangian. Second, it is the most highly
tuned of the Standard Model parameters, being $10^{-34}$ of its ``natural" value.
(The next most tuned parameter is $\overline{\theta}$, which is less than $10^{-9}$ of its natural
value.) Third, the smallness of $\mu^2$ is so far the
most intractable of the naturalness problems of the Standard Model. Various plausible mechanisms
of a more conventional sort have been proposed for
explaining the smallness of other Standard Model parameters. (For example, the smallness
of $\overline{\theta}$ can be explained by the Peccei-Quinn mechanism \cite{pq} or by spontaneously broken
CP or P \cite{sbcp}; and various symmetry schemes have been proposed to explain the smallness of the Yukawa couplings
of the light quarks and leptons). By contrast,
attempts to explain the smallness of the weak scale by technicolor or similar ideas are plagued by
a variety of well-known difficulties. And low energy supersymmetry does not by itself explain the magnitude of
the weak scale, though it protects it from radiative correction.

Since the focus of Ref. 1 was on the mass parameter of the Higgs field, the question naturally arises
how things would be different if there were two or more Higgs doublets. Two Higgs
doublets appear in a wide variety of theoretical contexts: in theories with supersymmetry; in Peccei-Quinn models;
and in many grand unified models (such as $SO(10)$, if the ${\bf 10}$ of Higgs fields is assumed to be complex).
The simplest possibility, which is typical of the above-mentioned scenarios, and which
avoids problems of Higgs-mediated flavor-changing processes, is that one doublet ($H_u$) couples to
the up-type quarks, while the other doublet ($H_d$) couples to the down-type quarks and charged leptons \cite{nfc}.
It is this situation we will analyze in this paper. Thus, at high scales we have

\begin{equation}
\begin{array}{ccl}
{\cal L}_{Yuk}&  = & (Y_u \overline{u} u + Y_c \overline{c} c + Y_t \overline{t} t) H_u \\ & & \\
& + & (Y_d \overline{d} d + Y_s \overline{s} s + Y_b \overline{b} b) H_d +
(Y_e \overline{e} e + Y_{\mu} \overline{\mu} \mu + Y_{\tau} \overline{\tau} \tau) H_d.
\end{array}
\end{equation}

\noindent
With two Higgs doublets, there are several dimensionful Higgs-mass parameters that may vary among domains,
with the consequence that $\langle H_u \rangle = v_u$ and $\langle H_d \rangle = v_d$ may vary
independently.
This means that there is the possibility of explaining not only the magnitude of the weak scale, but
also the relative magnitudes of the masses of the up-type and down-type quarks.  In particular, we
shall see that an explanation of the size of $m_u/m_d$ and thus a partial explanation of $m_t/m_b$ emerges.
Much of our analysis would apply equally to models with or without low-energy supersymmetry, and
would apply in fact to any model where in effect both $v_u$ and $v_d$ vary among domains.
Thus, the ``anthropic" considerations given
in this paper are not proposed as an alternative to low-energy supersymmetry.  Indeed, they may
shed light eventually on supersymmetry-breaking and the $\mu$ problem.  However, in this paper we will
consider for purposes of concreteness a non-supersymmetric two-Higgs-doublet model.

In such a model, the parameter $\mu^2$ is replaced by a two-by-two Hermitian mass matrix
$M^2_{ij}$, $i,j = u,d$, with four real parameters. To make the weak-scale small, still only one tuning is required,
namely of the determinant of $M^2_{ij}$, since only one Higgs doublet needs to remain light to break
$SU(2)_L \times U(1)_Y$. The other doublet has no reason to be light, and in fact would be expected to be
of superlarge mass.  The doublet that is light (whose mass-squared we shall call $\mu^2$)
is a linear combination of $H_u$ and $H_d$:
$H_{<} = \sin \beta H_u + \cos \beta \tilde{H}_d$, where
$\tilde{H}_d \equiv i \tau_2 H_d^*$.  The mixing angle $\beta$ enters into the relative
magnitudes of the up-quark masses and down-quark masses.  In particular, the ratios $m_u/m_d$
and $m_t/m_b$ are proportional to $\tan \beta$. If $M^2_{ij}$ varies in the landscape (or ``scans"),
then so does $\tan \beta$.
We shall see that ``anthropic" considerations strongly favor the observation of values of
$m_u/m_d$ that are of order (but somewhat smaller than) 1 and thus of values of $m_t/m_b$
much larger than 1 (assuming, as we shall, that quark and lepton Yukawa couplings are fixed, i.e. do not vary
among domains).

It is noteworthy that in published models or scenarios that do not invoke
``anthropic" or ``landscape" considerations the relations $m_u/m_d \sim 1$ and
$m_t/m_b \gg 1$ are generally not accounted for but are simply assumed to hold.
(However, there have been a few attempts to explain $m_u \sim m_d$ in more
conventional ways, such as \cite{bd}.)
It seems that anthropic and non-anthropic explanations are somewhat
complementary: those relationships
that seem most resistant to conventional dynamical or symmetry explanations (which
would include not only $m_u/m_d \sim 1$, but also $M_{weak}/M_{P \ell} \sim M_{strong}/M_{P \ell}$ and
$\Lambda/M_{P \ell}^4 \sim 10^{-120}$) are also those that are most straightforwardly
accounted for ``anthropically".

In this paper, as in Ref. 1, we assume that the dimensionless parameters of the Lagrangian are fixed,
i.e. do not vary among domains.  (More precisely, we assume that their values at some high scale are
fixed, since their low-scale values will be slightly influenced through renormalization group running by the values
of $v_u$ and $v_d$.)  If this is assumed, then the inter-family ratios of Yukawa couplings are known,
e.g. the ratios of $Y_u/Y_t$, $Y_c/Y_t$, $Y_d/Y_b$, and $Y_s/Y_b$.  We will also assume
that the Yukawa couplings of the third family are all of order 1.  This is plausible, given that
schemes that unify all the fermions of a family, such as $SO(10)$, relate these three couplings; and the simplest
such schemes give $Y_t = Y_b = Y_{\tau}$.

The relative values of the Yukawa couplings of the up-type and
down-type quarks are shown in a log plot in Fig 1.
\begin{figure}[h]
\begin{center}
\includegraphics{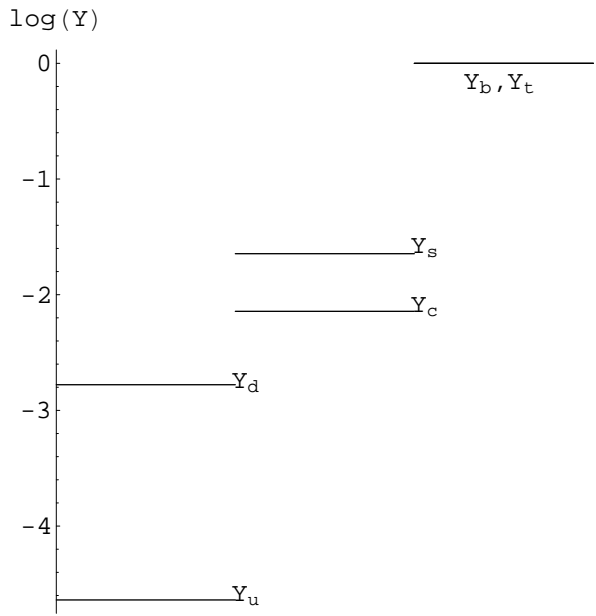}
\caption{The Yukawa couplings of the quarks if $Y_b \simeq Y_t \simeq 1$. The up-type
quarks have a much steeper family hierarchy. The scale on the left is $\log_{10} Y_i$.}
\label{Fig1}
\end{center}
\end{figure}

Since the mass hierarchy among the up-type
quarks is much steeper than that among the down-type quarks, normalizing
$Y_t \simeq Y_b$ implies that $Y_d \gg Y_u$. That is,

\begin{equation}
\left( \frac{Y_d}{Y_u} \right)  \equiv k =
\left( \frac{m_d}{m_u} \frac{m_t}{m_b} \right) \left( \frac{Y_b}{Y_t} \right) \noindent
\simeq 80 \left( \frac{Y_b}{Y_t} \right) \simeq 80.
\end{equation}

\noindent In the scenario we are discussing, the relative
normalization of the up-type masses and down-type masses is free to
``scan", i.e. vary among domains. The consequence of this, as we
shall see, is that $m_d/m_u$ is anthropically selected to be close
to but somewhat larger than 1, and therefore $m_t/m_b$ ends up (in
anthropically viable domains) being much larger than 1, as shown in
Fig 2.
\begin{figure}[h]
\begin{center}
\includegraphics{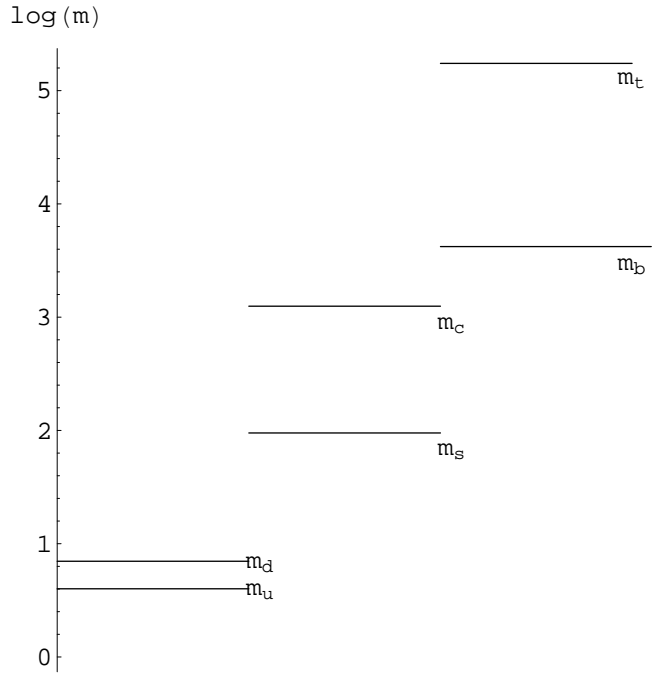}
\caption{If $v_u/v_d$ is anthropically set so $m_u/m_d \sim 1$,
then $m_t \gg m_b$ due to the steeper hierarchy of the up-type
quarks. The scale on the left is $\log_{10} (m_i/{\rm MeV})$.}
\label{Fig2}
\end{center}

\end{figure}
It is in this sense that one can anthropically ``explain" the
largeness of $m_t/m_b$.  Of course, this assumes the pattern of
interfamily mass ratios shown in Fig 1, i.e. the greater steepness
of the hierarchy for up-type quarks than for down-type quarks. There
exist many interesting and plausible theoretical schemes for
explaining that pattern. One example is the model of \cite{bb96},
where the hierarchy among the up-type quarks is quadratic and the
hierarchy among the down-type quarks is linear in the same small
parameters.

In this paper, we shall call domains of the universe
where organic life based on chemistry is possible
``viable", and we shall call the values of parameters in such domains
``anthropically allowed".
In section 2 of this paper, we shall examine the consequences for ``viability" of different
values of $m_u$ and $m_d$ and graphically
display the anthropically allowed and forbidden regions of
the $\ln m_u$-$\ln m_d$ plane.
In section 3, we shall look at the probability distributions of
$(m_u,m_d)$ that arise from certain simple assumptions about the Higgs mass matrix
$M^2_{ij}$. In section 4 we will use the probability distributions derived in section
3 to analyze the $\mu^2 >0$ region of parameter space in light of the existence of
dark energy and the anthropic requirement that it be fine-tuned to be small.

\section{Physics in the $\ln m_u-\ln m_d$ Plane}

In this section we will discuss anthropic constraints on $m_u$ and $m_d$. Most of
these constraints have been discussed in slightly different contexts in
\cite{abds} and \cite{hogan}, and much of the discussion in this section makes
use of their results.

The vacuum expectation $v$ that breaks the weak interactions and gives $u$ and $d$ their
masses is that of the lighter Higgs doublet $H_{<}$, whose mass-squared parameter
we are calling $\mu^2$.
We suppose that $\sqrt{|\mu^2|}$ has a superlarge ``natural" scale $M_*$, which
is of order $M_{P \ell}$ (or $M_{GUT}$, it does not matter for our analysis which).
There are two qualitatively very different cases, $\mu^2 < 0$ and $\mu^2 > 0$.
If $\mu^2 < 0$, as in our domain,
then $v \sim \sqrt{|\mu^2|}$ and is naturally of order
$M_*$. In this case, smaller $v$ corresponds to more fine-tuning of $M_{ij}^2$.
On the other hand, if $\mu^2 > 0$, the breaking of the weak interactions is only induced by
the coupling of $H_{<}$ to the quarks, which acquire condensates through the
dynamical breaking of chiral symmetry by QCD, so that $v \sim f_{\pi}^3/\mu^2$. In
this case, the natural scale of $v$ is $f_{\pi}^3/M_*^2$, and {\it larger} $v$ corresponds
to more fine-tuning.

One sees that the maximum and minimum values of $v$ are different by
a factor of $(M_*/f_{\pi})^3$. In other words, if $M_* \sim M_{P
\ell}$, $v$ can range over 60 orders of magnitude.  In fact, $v_u$
and $v_d$ each span the same range. This enormous space of possible
values of $m_u$ and $m_d$ is shown in Fig. 3.
\begin{figure}[h]
\begin{center}
\includegraphics{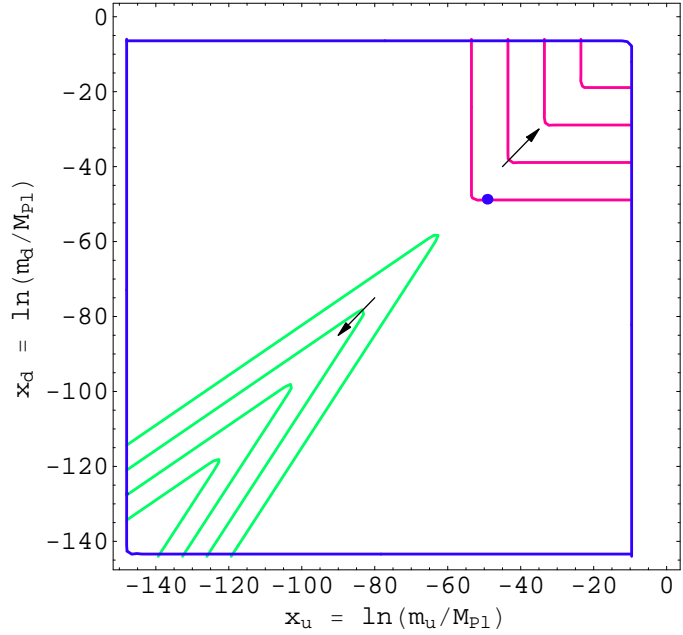}
\caption{The possible range of $m_i$, $i=u,d$, is from
$Y_i f_{\pi}^3/M_{P\ell}^2$ to $Y_i M_{P\ell}$. The blue dot shows the
observed values. The contours are of probability density in the landscape, red (green)
green for the $\mu^2 < 0$ ($>0$) region. Neighboring contours differ by $e^{20}$ in
probability density, increasing in the direction of the arrows.}
\label{Fig3}
\end{center}
\end{figure}
In that figure are also plotted the contours of equal probability density in the landscape 
(i.e. equal amount of fine-tuning of parameters),
which will be
derived in section 3 under the assumption that the parameters of $M_{ij}^2$ have a flat
probability distribution in the landscape in the interval $[- M_*^2, + M_*^2]$. The red curves are
contours of equal probability density in the $\mu^2 <0$ region, and the green curves
are contours of equal probability density in the $\mu^2 >0$ region. Neighboring contours differ in probability density by a factor of $e^{20}$, with the direction of increasing probability density indicated by the arrows.

The position of the``anthropically allowed" region in Fig. 3 in which our domain is
located is indicated by the blue dot. In fact, the viable region is so small that it
is smaller than the blue dot in Fig. 3.
Fig. 4 is a blow-up of that region of parameter space.

Since we are assuming that the Yukawa couplings of the quarks and leptons do not
vary among domains, and that both the down-type quarks and charged leptons get mass from
the same Higgs doublet $H_d$, the electron mass is given by $m_e = r m_d$, where
$r$ is a fixed ratio (up to logarithmic dependence on $v_u$ and $v_d$
through the running of parameters).  We take $m_{u0} = 4$ MeV and $m_{d0} = 7$ MeV, where the subscript
zero refers to the observed values of these parameters, i.e. their values in our domain
of the universe. Thus $r = 0.07$.

\subsection{The case of $\mu^2 < 0$}

First, we will discuss the case of negative $\mu^2$.  If $\mu^2$ is negative and smaller
in absolute value than in our domain, that involves more fine-tuning and is thus a less
probable region of parameter space.  Our interest is therefore primarily in $\mu^2$ negative
and greater than (in absolute value) or comparable to its value in our domain.  That
corresponds to values of
$m_u$ and $m_d$ greater than or comparable to their observed values.

If $m_d$ is sufficiently large, then all $d$ quarks will beta decay into $u$ quarks even
inside hadrons, so that the only stable hadron will be $\Delta^{++}$ ($uuu$).  ``Sufficiently
large" in this case means large enough to overcome the extra energy required to put three
quarks of the same flavor in a totally antisymmetric state, which energy we call $E_{\Delta N}$.
For $v_u$ and $v_d$ near the observed weak scale $v_0$, $E_{\Delta N} \sim 300$ MeV. (However, there
is a weak dependence of $E_{\Delta N}$ on $v_u$ and $v_d$ through the renormalization-group
running of $\alpha_s$, due to quark thresholds.)
The condition for all $d$ quarks to beta decay is $m_d \geq m_u + m_e + E_{\Delta N}$, or
$(1-r) m_d \geq m_u + E_{\Delta N}$. This is the region above the upper blue curve in Fig. 4,
which is a log-log plot. For $m_u$ and $m_d$ masses very large compared to
the QCD scale, this curve asymptotically approaches the straight line $(1-r) m_d = m_u$, which
is a line of slope 1 in Fig. 4.
\begin{figure}[h]
\begin{center}
\includegraphics{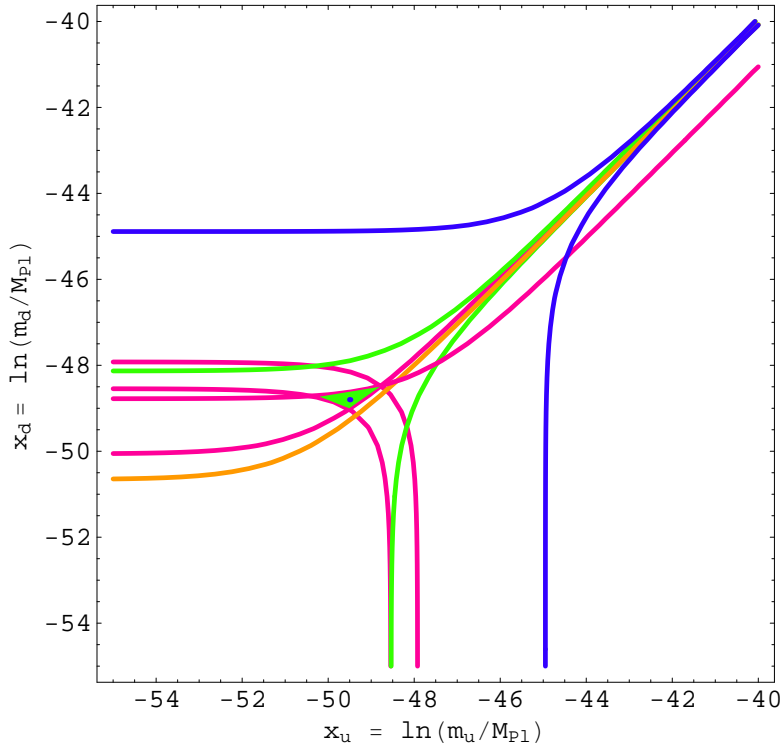}
\caption{A blowup of the region near our domain in Fig. 3. The curves
are anthropic bounds on $m_u$ and $m_d$ discussed in the text. The green shaded region
is potentially viable. The dot is our domain.}
\label{Fig4}
\end{center}
\end{figure}
For $m_u$ and $m_d$ small
compared to the QCD scale, the curve asymptotically approaches the horizontal line $m_d =
E_{\Delta N}/(1-r)$.

In the region where all the $d$ quarks decay, not only will the $\Delta^{++}$ be the only stable
hadron, it will surely also be the only stable nucleus. (In this regime, $m_d$ is large
compared to the QCD scale, so that the pions are not pseudo-goldstone particles and the range
of the nuclear potential is shorter than in nuclei of our own domain. Moreover, there is a strong Coulomb
repulsion between two $\Delta^{++}$.) These $\Delta^{++}$ will clothe themselves with a pair of
electrons, and act chemically like helium. With the only element being helium-like, it is extremely
implausible that life based on chemistry could exist.

On the other hand, if $m_u$ is sufficiently large compared to $m_d$, then all the $u$ quarks
will decay to $d$, and one will have a domain in which the only stable hadron and nucleus
will be a $\Delta^-$ ($ddd$). These will clothe themselves with positrons (which will presumably be more
numerous than electrons in such a domain, due to the beta decay of all the $u$ quarks),
and thus act chemically like hydrogen. An all-$\Delta^-$ universe is also extremely unlikely
to give rise to chemistry-based life. The all-$\Delta^-$ universe corresponds to
$m_u \geq m_d(1+r) + E_{\Delta N}$, which is the region below the lower blue curve in Fig 4.
(For large quark masses this curve
it is asymptotic to the line $(1+r) m_d = m_u$, which is a line of slope 1 in Fig. 4.)

Thus we can rule out almost all of the
three quadrants in Fig. 4 where either $m_u$ or
$m_d$ is large compared to its observed value. There is a thin ribbon between the two
blue curves where both $m_u$ and $m_d$ are large compared to the QCD scale
and $(1-r) m_d \leq m_u \leq (1+r) m_d$, where neither $d \rightarrow
u + e^- + \overline{\nu}_e$ nor $u \rightarrow d + e^+ + \nu_e$ can happen. However,
this region is unlikely to be viable, since $A=2$ nuclei are not bound there, as
shall be seen.

More of the parameter space is ruled out as viable if we take
into account the conditions that both neutrons and protons should be stable inside
nuclei. If $(A,Z)$ denotes a nucleus of mass $A$ and atomic number $Z$, then
the process $(A,Z) \rightarrow (A-1,Z) + p + e^- + \overline{\nu}_e$ is energetically
allowed if $m_d - m_u - m_e - \Delta m_{em} > B$, where $B$ is the binding energy
of the proton in the nucleus, and $\Delta m_{em}$ is the electromagnetic
contribution to the proton-neutron mass difference. Following
\cite{abds}, we take $\Delta m_{em} = 1.7$ MeV, and we take 10 MeV as a typical
value of $B$. (Of course, for very large values of $v_u$ and $v_d$, the values of $\Delta m_{em}$ and
$B$ would be somewhat different. However, in that limit $\Delta m_{em}$ and $B$ are
are negligible compared to
$m_u$ and $m_d$ anyway, so the shape of the curves is little affected.)
 This gives the condition $(1-r) m_d \geq m_u + 11.7$ MeV. If this is
satisfied, all neutrons beta decay, even inside nuclei. Since
stable nuclei having $A>1$ cannot be made up only of protons \cite{abds},
the only stable nucleus in such domains and therefore the only chemical
element would be $^1$H.  It is unlikely that chemical life can exist in an
such ``all hydrogen" domains. Thus viability requires that one has
$(1-r) m_d \leq m_u + 11.7$ MeV, which rules out the region above the upper green curve in
Fig. 4. Another constraint comes from the requirement that protons should be
stable in nuclei. If $m_u - m_d - m_e + \Delta m_{em} > B$,
then protons even in nuclei would be able to decay through the process $(A,Z) \rightarrow
(A-1, Z-1) + n + e^+ + \nu_e$. Viability thus requires
$m_u \leq (1+r) m_d + 8.3$ MeV, which rules out the region below the lower green curve in Fig. 4.

A stronger bound comes from the requirement that isolated protons are stable against
$p \rightarrow n + e^+ + \nu_e$. If they were not, there would be no $^1$H in such domains.
Since almost all organic molecules (with a few exceptions, such as CO$_2$) contain
hydrogen,
a no-hydrogen domain would presumably have dim prospects for life based on chemistry.
This rules out the region below the curve $m_d (1+r) = m_u + \Delta m_{em}$, shown as
the orange curve in Fig. 4.  A slightly stronger bound comes from the condition that
the proton in the nucleus of hydrogen not convert to a neutron by electron capture
$p + e^- \rightarrow n + \nu_e$, which rules out the region below the curve
$m_d (1-r) = m_u + \Delta m_{em}$, shown as the lower red,
concave-up curve in Fig. 4.

If the reaction $p + p \rightarrow D + e^+ + \nu_e$ in the sun is to be exothermic
one must have $m_d (1+r) \leq
m_u + B_D + \Delta m_{em}$, where $B_D = 2.2$ MeV is the binding energy of the deuteron.
A very similar condition comes from requiring that the deuteron be stable against the weak decay
$D \rightarrow p + p + e^- + \overline{\nu}_e$, namely
$m_d(1-r) \leq m_u + B_D + \Delta m_{em}$. In \cite{abds} the binding energy
of the deuteron was parameterized by

\begin{equation}
B_D = 2.2 - a \left( \frac{m_u + m_d}{m_{u0} + m_{d0}} - 1 \right),
\end{equation}

\noindent
where energies are in MeV and the parameter $a$ is very sensitive to the model of the
internucleon potential. Using two different models, \cite{abds} obtained
$a = 5.5$ MeV and $a = 1.3$ MeV. Using Eq. (3), the condition for the deuteron
to be stable against weak decay becomes
$(0.93 - a/11) m_d - (1 - a/11) m_u \leq (3.9 + a)$.  For $a = 5.5$ MeV, this rules out the
region above the upper, red, concave-up curve in Fig. 4.
As noted in \cite{abds}, it may be possible for
the hydrogen burning to occur in stars and for nucleosynthesis to proceed even
if the deuteron is unstable to weak decay, if its lifetime is sufficiently long.
Thus, a more clear-cut constraint comes from the condition that the deuteron be stable
against the (much faster) strong decay $D \rightarrow p + n$.
The condition for this is simply
that $B_D$ be positive.  Using $a = 5.5$ MeV, this gives
$m_u + m_d \leq 15.4$ MeV, which rules out the region above the upper, red, concave-down curve
in Fig. 4.

If the quark masses are too small, then the pion mass becomes small, and the nuclear force
becomes sufficiently long-range to bind the ``diproton", i.e. the $^2$He nucleus, as pointed out
by Dyson long ago \cite{dyson}. This would allow
hydrogen to burn in stars by the reaction $p + p \rightarrow ^2$He, which is very fast as it does not
involve the weak interactions.  Hydrogen-burning stars would have lifetimes orders of
magnitude shorter than the billions of years presumably required for complex organisms
to evolve. In \cite{abds,hogan} it was estimated that if $m_u + m_d < 0.75(m_{u0} + m_{d0}) =
8.25$ MeV, the diproton is bound. This rules out the region below the lower, red, concave-down curve
in Fig. 4.

Altogether, then, we see that the viable region is squeezed down to the
remarkably small shaded green area
bounded by the four red curves in Fig. 4. The actual value
of $(m_u, m_d)$ observed in our domain
is right in the middle of it, as shown in Fig. 4. (For perspective,
recall that this shaded green region is
smaller than the dot in Fig. 3, which shows the whole parameter space.)

\subsection{Exotic domains with light $c$ quarks}

Exotic possibilities open up if either $m_c$ and $m_s$ is small.
First we shall consider the situation where $v_u$ is so small that $m_c$ becomes
comparable to or smaller than $m_d$. (Of course, since the Yukawa coupling ratios are
fixed, $m_u$ becomes about two orders of magnitude less than $m_d$.)
We shall call these ``light-$c$-quark domains".
In such domains the light quarks are
$d$, $u$ and $c$, and there is a baryon octet made up of valence quarks of these three
flavors. This octet, shown in Table I, contains baryons of
charge 0, +1, and +2, but none with negative
charges. Therefore these domains are similar to ours in the following
respects: (1) All nuclei have $Z > 0$. (2) If a nucleus has large $A$, then its charge $Z$ is
also large, since Fermi energy tends to equalize the number of baryons of different
species. (3) For large $Z$ nuclei, Coulomb energy becomes important and gives an upper bound
to the size of stable nuclei.

\vspace{0.4cm}

\begin{displaymath}
\begin{array}{ccccc}
& {\bf \Xi^+_{cc}} & & {\bf \Sigma^0_c} & \\
& (ccd) & & (cdd) & \\ & & & & \\
 {\bf \Xi^{++}_{cc}} & & {\bf \Sigma^+_c}, {\Lambda^+_c} & & {\bf n} \\
(ccu) & & (cud) & & (udd) \\ & & & & \\
& {\bf \Sigma^{++}_c} & & {\bf p} & \\
& (cuu) & & (uud) & \end{array}
\end{displaymath}

\begin{center} {\bf Table I}
\end{center}

\vspace{0.4cm}

In the light-$c$-quark part of parameter space one can distinguish three regions,
which are shown in Fig. 5: (A) For large enough $m_c$,
the lightest baryons in the octet are those without valence $c$ quarks, namely $p$ and $n$.
For nuclei composed of $p$ and $n$ the analysis of the previous subsection applies.
Consequently, case (A) does not give any anthropically allowed regions of parameter
space except those we have already discussed.
(B) For small enough $m_c$, the lightest baryons in the octet are those with only
valence $c$ and $u$ quarks, namely $\Sigma^{++}_c$ and $\Xi^{++}_{cc}$ (i.e. $uuc$ and
$ucc$, respectively). As we shall see, domains of this sort seem unlikely
to be viable. (C) For intermediate values of $m_c$, the lightest baryons in the
octet are $p$ and $\Sigma^{++}_c$. Some domains of this sort may be viable.

The sufficient condition for case (A) is that $m_n < m_{\Sigma^{++}_c}$, which gives
$2 m_d < m_c + m_u + \Delta m'_{em}$, where the last term is the
electromagnetic contribution to the splitting of  $\Sigma^{++}_c$ and $n$. We may neglect $m_u$
in this expression, since it is less than a percent of $m_c$. We take $\Delta m'_{em}
= 4 \Delta m_{em} = 6.8$ MeV. (Note that the observed splitting between $\Sigma^{++}_c$ and
$\Sigma^0_c$ is 0.3 MeV, so that if the quark mass contribution is $2(m_u - m_d) = -6$ MeV,
it implies an electromagnetic contribution of 6.3 MeV. Of course, the wavefunctions for
the charmed baryons would be different in the light-$c$-quark domains.)
The condition is then $m_c > 2 m_d - 6.8$ MeV.
This is the region to the right of the lower red curve in Fig. 5.
\begin{figure}[h]
\begin{center}
\includegraphics{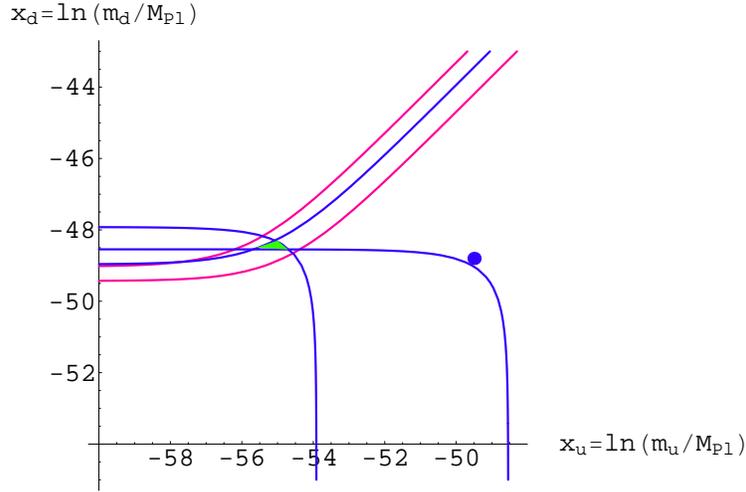}
\caption{The ``light-$c$-quark" region of parameter space. The green shaded region
may be viable. The dot is our domain.}
\label{Fig5}
\end{center}
\end{figure}
In this region,
the analysis we gave in section 2 applies, and one expects that only anthropically allowed
region is that shown in Fig. 4, which includes our domain (indicated by the blue dot in Fig.
5).

The condition for case (B) is that $m_{\Xi^{++}_{cc}} < m_p$, which gives
$2 m_c + 3 \Delta m_{em} < m_d + m_u$, or approximately
$m_c < m_d/2 - 2.5$ MeV. This is the region to the left of the upper red curve
in Fig. 5.  This region seems unlikely to be viable since all stable nuclei would have
$Z \geq 2$. (It is easily
seen that an isolated $p$ will beta decay to $\Sigma^{++}_c$, as will a $p$
bound with neutral baryons.) Consequently, there would be no chemical analogue
of hydrogen.

The most interesting case is (C), which lies in between the two red curves in
Fig. 5. Here the lightest two baryons are $p$ and $\Sigma^{++}_c$.
In such a domain, the
only $Z = 1$ nucleus consists of a single proton, i.e. $^1$H. In order for
hydrogen to be stable, it is required that the decay
$p \rightarrow \Sigma_c^{++} + e^- + \overline{\nu}_e$ be energetically
forbidden, which gives the condition
$m_c > m_d (1 - r) - 3 \Delta m_{em}$. This rules out the region above the blue,
concave-up curve in Fig. 5.
A second condition is that diprotons not be bound, as otherwise hydrogen-burning
stars would burn too fast, as noted above. This gives the bound that
$m_d + m_u > 8.25$ MeV, as given previously.
This rules out the region below the nearly horizontal blue line that barely misses our
domain in Fig. 5.

A third condition is that the $A = 2$ nucleus composed of a $p$ and a
$\Sigma_c^{++}$ be bound. Let us call this analogue of the deuteron
$\tilde{D}$. The potential
between the $p$ and the $\Sigma_c^{++}$
at large distances is given by one-$D$-meson exchange rather than
one-pion-exchange, and is thus controlled by $m_c + m_d$ rather
than by $m_u + m_d$. Using Eq. (3), the binding energy of $\tilde{D}$ can
be written $B_{\tilde{D}} = 2.2 - E_c - a [(m_c + m_d)/(m_{u0} + m_{d0}) - 1]$,
where $E_c$ is the Coulomb interaction energy of the baryons in the $\tilde{D}$.
The condition that $\tilde{D}$ be bound is then
$m_c + m_d < 11 [(2.2 - E_c)/a +1)$, where energies are in MeV. If
$a = 5.5$ MeV and $E_c = 2$ MeV, this rules out the region above the blue, concave-down
curve in Fig. 5.

The combination of the three constraints leaves only the very small triangular
region shaded green in Fig. 5. In fact, since the three curves depend on quantities
that are not well known, such as $a$, $E_c$, $m_{u0}$, $m_{d0}$, it is not
really clear that {\it any} area is actually enclosed by them. Furthermore, even
if there is, it does not necessarily mean that this area is an anthropically allowed
region of parameter space. It only means that three of the more obvious
disasters for viability do not happen. One other potential problem for
viability is that nucleosynthesis may be inhibited in such domains because
of large Coulomb barriers, due to the fact that nuclei are composed of charge-1
and charge-2 baryons. A much more serious potential problem for viability is that
$G_F$ is about four orders of magnitude larger in this region
of parameter space than it is in our domain. This would make the ``$pp$ reaction"
$p + p \rightarrow \tilde{D} + e^- + \overline{\nu}_e$ proceed in
hydrogen-burning stars very much faster than the ordinary $pp$ reaction does
in our domain. That would presumably be disastrous for
the same reason that the diproton being bound would be.

As we shall see in section 3, where we discuss probability distributions
in parameter space under some simple and reasonable assumptions,
these potentially viable light-$c$-quark domains occur
less commonly in the universe than domains like our own. Therefore,
unless a typical domain of this sort is {\it more} viable than ours, the existence
of such domains should not affect the
anthropic explanation of the observed values of $m_u$ and $m_d$ that we are
exploring here.

\subsection{Exotic domains with light $s$ quarks}

We now consider the situation where $v_d$ is so small that the mass of the $s$
quark is comparable to or smaller than $m_u$. (Of course, then $m_d \ll m_u$, since
we are assuming that the ratios of Yukawa coupling are fixed.) We will call these
``light-$s$-quark domains".
In such domains, the lightest three quarks are $d$, $s$, and $u$, as in our domain,
but the corresponding flavor $SU(3)$ is a very good symmetry, since $m_s$ is small.
Consequently, there is the familiar baryon octet, shown in Table II, but it is more
nearly degenerate than in our domain.

\vspace{0.4cm}

\begin{displaymath}
\begin{array}{ccccc}
& {\bf \Xi^0} & & {\bf \Sigma^+} & \\
& (uss) & & (uus) & \\ & & & & \\
 {\bf \Xi^-} & & {\bf \Sigma^0}, {\Lambda^0} & & {\bf p} \\
(dss) & & (uds) & & (uud) \\ & & & & \\
& {\bf \Sigma^-} & & {\bf  n} & \\
& (dds) & & (udd) & \end{array}
\end{displaymath}

\begin{center} {\bf Table II}
\end{center}

\vspace{0.4cm}

There are three cases to consider for light-$s$-quark domains (just as there were for
light-$c$-quark domains). These are shown in Fig. 6. (A) For large enough $m_s$,
the lightest two baryons are the two non-strange baryons $p$ and $n$, as in our domain.
The condition for this is that $m_{\Sigma^-} > m_p$, which gives
$m_s + m_d > 2 m_u$. This is the region above the upper red curve
in Fig. 6. (B) For small enough
$m_s$, $u$ is heavier than both $s$ and $d$ and the lightest two baryons are
$\Sigma^-$ and $\Xi^-$, which have no valence $u$ quarks.
The condition for case (B) is that $m_{\Xi^-} < m_n$, which gives
$m_u + m_d  > 2 m_s + \Delta m_{em}$. This is the region below the lower red
curve in Fig. 6.
(C) For intermediate $m_s$, the lightest two baryons would be $n$ and $\Sigma^-$.
This is the region between the two red curves in Fig. 6.

For case (A), where the lightest baryons are $p$ and $n$, as in our domain, the
analysis of viability we presented earlier applies, and so the only
anthropically allowed region is that shown in Fig. 4, which includes our domain
(indicated by the blue dot in Fig. 6).
\begin{figure}[h]
\begin{center}
\includegraphics{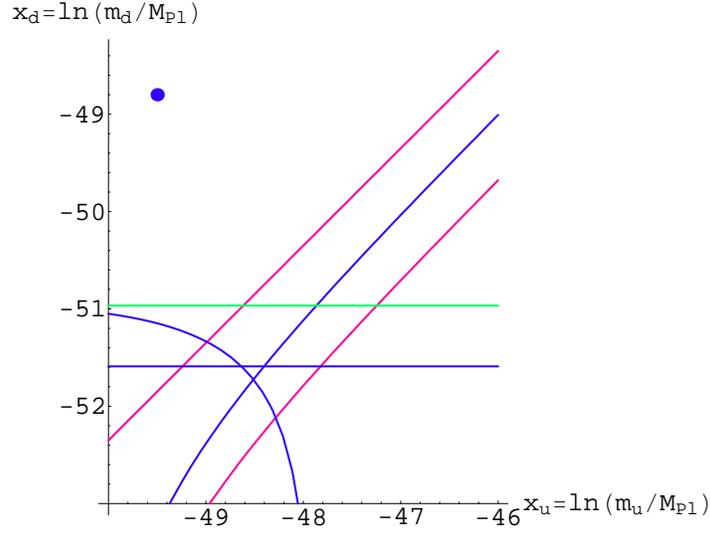}
\caption{The ``light-$s$-quark region of parameter space. Some of the region below the
green line and to the right of the rightmost red curve may be viable. The dot is
our domain.}
\label{Fig6}
\end{center}
\end{figure}
Case (C) is interesting because it might lead to domains in which the $\Sigma^-$
plays the role that the proton plays in our domain.
For example, a ``hydrogen atom" in such a domain would consist of a $\Sigma^-$ and an $e^+$.
(One would expect antileptons to predominate in such domains because of charge
neutrality.) There are at least three constraints that $m_u$ and $m_s$ must satisfy in
case (C).

First, the $\Sigma^-$ in the nucleus of the ``hydrogen" atom must be stable against
beta decay into $n + e^- + \overline{\nu}_e$. This gives the constraint
$m_u > (1 - 0.05 r) m_s + \Delta m_{em}$. This rules out the region to the left of
the blue curve that is concave to the right in Fig. 6.

The second constraint is that the $n$ and $\Sigma^-$ must be able to bind to make
the analogue of the deuteron.
The long-distance part of the potential between these nucleons is controlled by
the mass of the kaon, and thus by $m_u + m_s$. Thus the condition for the
$n$ and $\Sigma^-$ to bind is $m_u + m_s < 15.4$ MeV, if $a = 5.5$ MeV.
This rules out the region above the blue curve that is concave to the left in
Fig. 6. The third constraint is that the di-$\Sigma^-$ not be bound (the analogue of the
diproton not being bound in our domain). This gives $m_s + m_d > 8.25$ MeV, which rules out
the region above the nearly horizontal blue line in Fig. 6.
Altogether, there is no allowed region left for case (C). However, this depends on the
value of $a$, and other parameters that are not precisely known. If $a$ is less
than about 3 MeV, there may be a small region that satisfies the three constraints.
Of course, even if there were, that hardly proves that such a region is anthropically
allowed; it merely means that some obvious disasters for viability are avoided.

We turn now to case (B). In this case the lightest baryons are $\Sigma^-$
and $\Xi^-$, and $\Sigma^-$ is the lighter of the two. The only obvious constraint
is that the analogue of deuterium (i.e. a bound state of $\Sigma^-$ and $\Xi^-$)
be allowed to form. For $a = 5.5$ MeV, this gives the constraint $m_s + m_d =
1.05 m_s < 15.4$ MeV. This rules out the region above the horizontal green curve in
Fig. 6.

There may be some anthropic upper bound on $m_u$, coming from
the fact that very large $m_u$ corresponds to very small $G_F$; however, that is not
clear. If $v_u$ is superlarge and therefore
$G_F \sim v^{-2}$ is supersmall, rather strange situations are possible.
Even though $G_F$ is supersmall, the decay rates of $u$, $c$, and $t$ quarks,
being of order $G_F^2 m_{u,c,t}^5 \sim v_u$ will be very fast,
and the only quarks around will be $d$, $s$, and $b$.
The amplitudes for transitions of $d$, $s$, and $b$ into
each other will be suppressed by a factor of $v_u^{-2}$ and
therefore, in effect, there will be separate conservation
of $d$-number, $s$-number, and $b$-number.  (Sphaleron
processes will, of course also be highly suppressed.)
The story is similar for the leptons; there will be
separate conservation of $L_e$, $L_{\mu}$, and $L_{\tau}$.

In this case, much depends on the details of primordial
lepto/baryogenesis.  It is possible that lepto/baryogenesis
leads to asymmetries for $d$, $s$, and $b$ quarks that
differ in sign.  For example, there could be an
excess of $d$ over $\overline{d}$, but of $\overline{s}$ over
$s$ and $\overline{b}$ over $b$.  In that case, ``ordinary"
hadronic matter would be made up of $d$, $\overline{s}$,
and $\overline{b}$.  In particular, there would be stable
$K^0$, $B^0$ mesons,  $\Sigma^-$ baryons, and positively
charged antibaryons consisting of valence $\overline{s}$
and $\overline{b}$.  These could capture leptons of positive
or negative charge (as the case may be) and make atoms.
In the same way, there could be asymmetries of different
sign for different flavors of lepton.  For instance, there
could be an excess of $e^-$ and $\mu^-$ but of $\tau^+$.
Thus there could be atoms consisting of an $e^-$ orbiting
a $\tau^+$. Clearly, all sorts of rich and strange
possibilities for chemistry might exist.  And if the
chemistry is rich, then it is plausible that life might be possible.
However, in order to have any nuclei with $Z>1$, one still needs
two baryons to be able to bind. There is still, therefore, an
anthropic upper bound on $m_d + m_s$ for case (B), as discussed above.
As is clear from the shape of the probability contours shown in
Fig. 3 and derived in section 3, this will put an upper bound on the
probability density of viable small-$s$-quark domains.

In sum, there are regions of parameter space with $m_s$ of order a few MeV,
and in particular case (B), where we have found no very obvious convincing
argument against viability. There may also be a small anthropically
allowed region where $m_c$ is of order a few MeV.
As we will see in section 3,
for certain plausible assumptions about the probability distribution in the
$\ln m_u - \ln m_d$ plane, the possibility of such exotic viable
light-$s$-quark or light-$c$-quark domains
would not affect the anthropic explanation of the observed values
of $m_u$ and $m_d$ being explored in this paper.

\subsection{The case of $\mu^2 >0$}

In domains where $\mu^2 > 0$ the vacuum expectation value of $H_<$ is of order
$f_{\pi}^3/\mu^2$ and therefore extremely small compared to its value in our domain.
This means, of course, that in these domains the quark and lepton masses are
correspondingly small.   (For example, if $\mu^2 \cong  + (100 {\rm GeV})^2$ and
$\tan \beta$ is the same as in our domain, then the electron mass is only about
$5 \times 10^{-4}$ eV.)    This has important consequences for life based on chemistry.
The reason is that the energies of typical atomic and molecular transitions are of
order, or smaller than, $\alpha^2 m_e$, and therefore so are the temperatures needed
for biological processes to occur.  (That is why in our domain the temperature needed
for life is about 300 K $ \cong 10^{-3} \alpha^2 m_{e0}$.)  It follows that before life
based on chemistry can appear the universe has to have cooled down to a temperature much
smaller than $\alpha^2 m_e$.

Because $m_e$ is so small in positive-$\mu^2$ domains, it was argued in \cite{abds}
that by the time the universe cools sufficiently for chemistry-based life to exist
in those domains various disasters might already have occurred, such as all baryons
having decayed away or all stars having burned out.  Those arguments depended
crucially on the assumption that universe expands as a power of $t$.  However,
it now appears that after dark energy comes to dominate the universe will
expand exponentially in $t$.  Once the expansion becomes exponential, it
does not take long for the universe to reach extremely small temperatures.
For example, consider a domain where $m_e = 5 \times 10^{-4}$ eV. If there
were no dark energy, it would take such a domain about $10^{20}$ years to
reach temperatures of order $10^{-3} \alpha^2 m_e$.  On the other hand,
if there were the observed amount of dark energy, it would take only about
$10^{11}$ years.  The difference is even more dramatic if $\mu^2$ is larger.
In a typical domain where $\mu^2 \cong (10^{10} {\rm GeV})^2$, it would take about
$10^{44}$ years to reach biochemical temperatures if there were no dark energy,
but only about $10^{13}$ years if there is the amount of dark energy we observe.

The existence of dark energy means that the specific astrophysical and cosmological
anthropic arguments used for the positive-$\mu^2$ case in \cite{abds} are not valid.   However,
there are other anthropic arguments based on astrophysical and cosmological considerations
that become possible precisely because of the existence of dark energy, as we will now see.
For the rest of the paper we will assume that the dark energy has $w = -1$ and will refer
to it as ``the cosmological constant".   Our arguments should not be qualitatively affected
if $w$ is not exactly $-1$.

It seems reasonable to suppose that life requires the existence of planets
upon which to evolve.  A planet is supported against gravitational collapse
by the electrons, and has therefore a typical density that goes as
$\rho_{planet} \sim A m_p (\alpha m_e/2)^3$, where $A$ is the average
number of baryons per nucleus.  A planet will be ripped apart by the
inflation caused the cosmological constant, unless

\begin{equation}
\rho_{\Lambda} < \rho_{planet} \sim A \alpha^3 m_e^3 m_p.
\end{equation}

\noindent
If $\rho_{\Lambda}$ is truly a constant of nature, so that it the same
in every domain, then its measured value of about $10^{-123} M_{P \ell}^4$
means that planets can exist only if $m_e > 6 \times 10^{-5}$ eV.  For
$\tan \beta$ having the same value as in our domain, this means that
$\mu^2$ has to be less than about (300 GeV)$^2$.  This is significantly stronger than the
bound given in Eq. (6) of \cite{abds} (a bound that came from not having baryons
decay away before the universe cooled to temperatures where chemical life could exist).

However, it is more interesting to suppose that $\rho_{\Lambda}$,
as well as the Higgs mass parameters, varies among domains.  Indeed,
this comports with the idea that all the dimensionful parameters
of the Lagrangian are determined anthropically.  In that case one has to
consider simultaneously the probability distributions of $\mu^2$, $\tan \beta$
and $\rho_{\Lambda}$.  This will be done in section 4.  It will be seen that
the viability of a significant part of the positive $\mu^2$ parameter space
is left unaffected by the ``planet argument".  However, most of that part
of parameter space can be anthropically ruled out by a second argument,
which is based on structure formation.

That second argument assumes that structure cannot begin to form until
recombination occurs.  If by the time of recombination the density of
the universe is already dominated by the cosmological constant, then
structure will never form.  The recombination temperature $T_{rec}$
is about $0.05 \alpha^2 m_e$, so the constraint is that structure cannot
form unless

\begin{equation}
\rho_{\Lambda} < \rho_B (t_{rec}) \sim m_p \eta_B T_{rec}^3 \sim 10^{-14} \alpha^6
m_e^3 m_p.
\end{equation}

\noindent
This appears to be a much more stringent constraint than the one in Eq. (4).
However, it is not clear that it can be used for $m_e$ smaller than about 20 eV.
The point is that, if
$m_e < 20$ eV, then when matter begins to dominate over
radiation (at $T \sim 1$ eV) the electrons will still be almost relativistic,
and the thermal number density of electrons and positrons will be larger than that of the baryons
by a factor of about $10^{10} e^{-m_e/T}$.
There will then be of order $10^{10}$ electrons and positrons for each baryon.
The charges of the baryons will be Debye screened, and it would seem that baryons
can already begin to condense at that point, since $\delta \rho_B/\rho_B$ would
not have a significant effect on the density of the leptons and photons.  We
will therefore only use Eq. (5) when $m_e > 20$ eV.  The anthropic bounds
coming from Eqs. (4) and (5) will be analyzed in section 4.

\section{Probability distributions in the $\ln m_u- \ln m_d$ plane}

Since anthropic explanations of relations among parameters are based
on probabilities, one must be able to make reasonable hypotheses about
probability distributions within the parameter space or ``landscape".
The parameters that we are assuming to vary among domains of the universe
are those in the Higgs mass-squared matrix $M^2_{ij}$. This matrix acts
in the space of $\left( \begin{array}{c} H_u \\ \tilde{H}_d \end{array} \right)$.
It is convenient to parameterize this matrix as follows

\begin{equation}
M^2 = \left( \begin{array}{cc} a_0 + a_3 & a_1 - i a_2 \\
a_1 + i a_2 & a_0 - a_3 \end{array} \right).
\end{equation}

\noindent
First, we will assume that the four real parameters $a_{\alpha}$ (which
have dimensions of mass squared) are ``naturally" superlarge
(either $O(M_{GUT}^2)$ or $O(M_{P \ell}^2)$)
and of the same order. Call their natural scale $M_*^2$. The simplest
assumption is that the four $a_{\alpha}$ are independent and that each has a flat
probability distribution in the interval $[-M_*^2, +M_*^2]$. (However, symmetry may
somewhat suppress
the off-diagonal elements of $M_{ij}^2$. We will discuss the effects of this later.)
We will take $a_0$
to be positive. (If it is negative and very large the weak interactions
are broken at very large scales, whereas if it is negative and small then
breaking the weak interactions at small scales would require that
$a_3$ also be small, which is an extra fine tuning.) It is convenient
to think of the parameters $a_1$, $a_2$, and $a_3$ as forming the components
of a vector $\vec{a}$ in a three-dimensional space.  As these three parameters have
been assumed all to
have flat probability distributions in the interval $[-M_*^2, +M_*^2]$,
the probability distribution is rotationally invariant in the
space of $\vec{a}$.

We shall denote the larger eigenvalue of $M^2_{ij}$ by $M^2_{>}$ and the smaller
eigenvalue by $\mu^2$.  (By our assumption about $a_0$, $M_{>}^2$ is positive.)
Since $M^2_{>}$ is generally much larger than $\mu^2$, the heavier Higgs doublet
decouples; so the parameter
$\mu^2$ is in effect just the $\mu^2$ of the Standard Model
and controls the breaking of the weak interactions. Because the cases of anthropic
interest are those where $|\mu^2| \ll M_{>}^2$, we may write

\begin{equation}
\mu^2 \cong \frac{\det M^2}{{\rm tr} M^2} = \frac{a_0^2 - \vec{a}^2}{2 a_0},
\end{equation}

\begin{equation}
\tan 2 \beta = \frac{2 |M^2_{ud}|}{M^2_{uu} - M^2_{dd}}
= \frac{\sqrt{a_1^2 + a_2^2}}{a_3} = \tan \theta_a,
\end{equation}

\noindent
where $\beta$ is the angle between the mass basis of the Higgs doublets
$(H_>, H_<)$ and the
original flavor basis $(H_u, \tilde{H}_d)$ (i.e. $H_{<} = e^{i \alpha} \sin \beta H_u +
\cos \beta \tilde{H}_d$ and
$\tan \beta = \left| \frac{v_u}{v_d} \right|$),
and $\theta_a$ is the angle between $\vec{a}$ and the $a_3$ axis.
There are two cases to consider, $\mu^2 <0$ and $\mu^2 > 0$.

\subsection{Probability distribution in the case of $\mu^2 < 0$}

In the case of negative $\mu^2$, the vacuum expectation value
that breaks the weak interactions is $v = \sqrt{|\mu^2|/\lambda} =
(2 \lambda)^{-1/2}$ $ \left| (a_0^2 - a^2)/a_0 \right|^{1/2}$, where
$a$ is the magnitude of $\vec{a}$, and $\lambda$ is the quartic self-coupling of
the lighter Higgs doublet.
Then we may write
$m_u = Y_u v_u = Y_u v \sin \beta \cong
(Y_u/\sqrt{2 \lambda})$ $ \sin \frac{\theta_a}{2} \sqrt{(a_0^2 - a^2)/a_0}$,
and similarly $m_d = Y_d v_d = Y_d v \cos \beta \cong (Y_d/\sqrt{2 \lambda})
\cos \frac{\theta_a}{2}$ $ \sqrt{(a_0^2 - a^2)/a_0}$.
Defining $K = Y_u/\sqrt{2 \lambda}$ and recalling the definition of $k$ in Eq. (2),
one has

\begin{equation}
\begin{array}{cl}
& m_u = K \sin \frac{\theta_a}{2} \sqrt{(a_0^2 - a^2)/a_0}, \\
m_d/k \equiv & \overline{m_d} = K \cos \frac{\theta_a}{2} \sqrt{(a_0^2 - a^2)/a_0}.
\end{array}
\end{equation}

\noindent
Since, by our assumptions, the probability distribution is rotationally invariant in
the space of the vector $\vec{a}$, one may write the probability distribution in
spherical coordinates $(a, \theta_a, \phi_a)$ in that space:

\begin{equation}
P(a_0, a, \theta_a, \phi_a) \; da_0 \; da \; d \theta_a \; d \phi_a
= N \; da_0 \; a^2 \; da \; \sin \theta_a \; d \theta_a \; d \phi_a,
\end{equation}

\noindent
where $N$ is the appropriate normalization factor. We would like to compute
the probability distribution in the space of $m_u$ and $m_d$.  Thus we write
$P(a_0, a, \theta_a) da_0 \; da \; d \theta_a \; d \phi_a = P(a_0, m_u, \overline{m_d})
da_0 \; d m_u \; d \overline{m_d} \; d \phi_a$, and compute the Jacobian for the
transformation of coordinates:

\begin{equation}
J = \left| \frac{\partial (a_0, m_u, \overline{m_d})}{\partial (a_0, a, \theta_a)}
\right| =  \frac{1}{2} K^2 (a/a_0).
\end{equation}

\noindent
This yields the result

\begin{equation}
P(a_0, m_u, \overline{m_d}) = 2 N \; K^{-2} \; a_0 \; a  \; \sin \theta_a.
\end{equation}

\noindent
>From Eq. (9), one has $a = \sqrt{a_0^2 - a_0 K^{-2} (m_u^2 + \overline{m_d}^2 )}$
and

\begin{equation}
\sin \theta_a = \frac{2 m_u \overline{m_d}}{ m_u^2 + \overline{m_d}^2}.
\end{equation}

\noindent
This yields

\begin{equation}
P(a_0, m_u, \overline{m_d}) = 4 N \; K^{-2} \; a_0 \;
\sqrt{a_0^2 - a_0 K^{-2} (m_u^2 + \overline{m_d}^2)}
\frac{m_u \overline{m_d}}{m_u^2 + \overline{m_d}^2}.
\end{equation}

\noindent
For the regions of parameter space that are anthropically
interesting (i.e. where life is likely to be possible) one
has $K^{-2} m_u^2, K^{-2} \overline{m_d}^2 \ll a_0 \sim M^2_*$, so that the
previous equation can be written
$P(a_0, m_u, \overline{m_d}) = 4 N K^{-2} a_0^2 m_u \overline{m_d}/
(m_u^2 + \overline{m_d}^2)$. One then simply integrates over
the variable $a_0$ to get

\begin{equation}
P(m_u, \overline{m_d}) \; dm_u \; d\overline{m_d}
= N' \frac{m_u \overline{m_d} dm_u d\overline{m_d}}
{m_u^2 + \overline{m_d}^2}, \;\; {\rm for} \;\; \mu^2 < 0.
\end{equation}

It is more interesting to express the probabilities in terms of
the logarithms of the quark masses.  Defining $x_u =
\ln(m_u/M_*)$, $x_d = \ln (m_d/M_*)$, and $\overline{x_d}
= \ln (\overline{m_d}/M_*) = x_d - \ln k$, one has

\begin{equation}
P(x_u, x_d) \; dx_u \; d x_d =
N^{\prime \prime} \frac{ e^{2 x_u + 2 \overline{x_d}}}
{e^{2 x_u} + e^{ 2 \overline{x_d}}} \; dx_u \; d\overline{x_d},
\;\; {\rm for} \;\; \mu^2 < 0.
\end{equation}

\noindent
The contours of constant probability given by this expression are
plotted as the concave-up curves in Fig. 3.  Note the shift represented
by the fact that $P$ depends on $\overline{x_d} = x_d - \ln k$, where $k \simeq 10^2$.
This shift means that domains where $m_d \cong m_{d0}$ and $k^{-1} m_{u0} <
m_u < m_{u0}$ are just as likely to occur as domains like ours.  The reason for
this is simple.  In our domain, one has $m_d \sim m_u$, even though $Y_d =
k Y_u \gg Y_u$. This means that $v_u/v_d = \tan \beta \simeq k \gg 1$, so that
$v = \sqrt{v_u^2 + v_d^2} \cong v_u$. Reducing
$m_u$ from its value in our domain ($m_{u0}$) while holding $m_d$ fixed,
means reducing $v_u$ while holding $v_d$ fixed. This has the effect of reducing
$v$ and thus $\mu^2$, which increases the fine tuning. However, it also has the effect
of making the ratio $v_u/v_d$ closer to one, which {\it reduces} the fine-tuning.
(The factor of $\sin \theta_a = \sin 2 \beta$ in $P$ prefers equal values of $v_u$ and $v_d$.)
These effects cancel, so that it does not ``cost probability" to move to
the left in Fig. 2 from the point representing our domain down to values $m_u \sim k^{-1} m_{u0}$.
Decreasing $m_u$ more than that does cost probability, however, because
it means making $v$ and $\mu^2$ more fine-tuned without any compensating effect
from $\tan \beta$, which indeed starts to move away again from 1 towards 0.

Thus, to explain why we don't observe a value of $m_u$ as small as $k^{-1} m_{u0}
\sim 10^{-2} m_{u0}$, one cannot argue that such domains
are less common. Rather, one has to argue that they are less viable.
In fact, we have presented such arguments in section 2.
What makes the region of parameter space with $m_u \ll m_{u0}$ less viable
is a combination of constraints coming from reactions in hydrogen-burning stars
like the sun, as one sees from Fig. 4.  If $m_u$ is even slightly smaller than
$m_{u0}$, it either makes the $pp$ reaction
$p + p \rightarrow D + e^+ + \nu_e$ endothermic (since protons become
lighter relative to neutrons), or it makes the diproton a bound state, allowing
the reaction $p + p \rightarrow$ $^2$He, or it has both effects.
Another effect that is probably deleterious to the chances of life is
that as $v_u$ gets smaller $v$ gets smaller, as mentioned, and so $G_F$ gets larger.
This would make the $pp$ reaction go much faster (since it involves the weak interaction),
presumably making stars like the sun burn much faster. However, we have not
calculated this effect, since the reaction $p + p \rightarrow$ $^2$He being
allowed is an even more drastic effect.

\subsection{Probability distribution in the case of $\mu^2 >0$}

If $\mu^2$ is positive, then the light Higgs doublet $H_{<}$ only acquires a vacuum
expectation value due to quark-antiquark condensation at $f_{\pi}$, the chiral-symmetry-breaking
scale of QCD. Specifically, the lighter Higgs doublet has the coupling
$H_{<} (Y_t \sin \beta \overline{t}t + Y_b \cos \beta \overline{b}b)$.  After QCD chiral symmetry
breaking this produces a linear term in the effective potential for $H_{<}$ that is
of the form $(Y_t \sin \beta + Y_b \cos \beta) f_{\pi}^3 H_{<}$.  Since we have assumed that
$Y_t \simeq Y_b \simeq 1$, the coefficient $(Y_t \sin \beta + Y_b \cos \beta)$ is a slowly varying
function of $\beta$ that is of order 1. We will denote this $\gamma(\beta)$.

Then instead of Eq. (9) one has

\begin{equation}
\begin{array}{cl}
& m_u = Y_u \gamma(\beta) f_{\pi}^3 \sin \frac{\theta_a}{2} \left( (a_0^2 - a^2)/2 a_0 \right)^{-1} \\
m_d/k \equiv & \overline{m_d} = Y_u \gamma(\beta) f_{\pi}^3 \cos \frac{\theta_a}{2} \left( (a_0^2 - a^2)/2 a_0 \right)^{-1}.
\end{array}
\end{equation}

\noindent
Thus, the Jacobian is given not by Eq. (11) but by

\begin{equation}
J = (Y_u \gamma(\beta) f_{\pi}^3)^2 4 a_0 a^2/(a_0^2 - a^2)^3.
\end{equation}

\noindent
There may be values of $\beta$ for which $\gamma(\beta) \ll 1$ due to an almost exact
cancelation between the $Y_t$ term
and the $Y_b$ term. However, that simply makes $m_u$ and $m_d$
small. Since smaller values of $m_u$ and $m_d$ are the the most probable ones anyway in the case
of $\mu^2 >0$, the possibility of such cancelations has virtually no effect on the probability
distribution we shall derive. Thus we will get qualitatively the right behavior for
$P(m_u, m_d)$ if we treat $\gamma$ as a constant, which we will now do for simplicity.
Then following the same steps as before one ends up with

\begin{equation}
P(m_u, \overline{m_d}) \; dm_u \; d\overline{m_d}
= \tilde{N}' \frac{m_u \overline{m_d} dm_u d\overline{m_d}}
{(m_u^2 + \overline{m_d}^2)^{5/2}}, \;\; {\rm for} \;\; \mu^2 > 0.
\end{equation}

\noindent
and, in terms of the variables $x_u$ and $\overline{x_d}$,

\begin{equation}
P(x_u, x_d) \; dx_u \; d x_d =
\tilde{N}^{\prime \prime} \frac{ e^{2 x_u + 2 \overline{x_d}}}
{(e^{2 x_u} + e^{ 2 \overline{x_d}})^{5/2}} \; dx_u \; d\overline{x_d},
\;\; {\rm for} \;\; \mu^2 > 0.
\end{equation}

The probability contours for both positive and negative $\mu^2$ that we have
just derived have been displayed in Fig. 3.
For negative $\mu^2$ the probabilities increase upward and to the left (i.e. towards larger
$m_u$ and $m_d$), whereas for positive $\mu^2$ the probabilities increase
downward and to the left (i.e. toward smaller $m_u$ and $m_d$). The directions
of increasing probability are shown by the arrows and the contours with equal
probability to our own domain are shown by the red curves. We see from this that
the potentially viable ``light-$s$ quark domains" and ``light-$c$-quark domains"
are less common than domains like our own.

In deriving the probability distributions given in Eqs. (16) and (20), we assumed that all
the elements of the Higgs mass matrix $M^2_{ij}$ have the same natural scale $M_*^2$.
However, the ``natural flavor conservation" \cite{nfc} pattern of Yukawa couplings
that we have assumed in Eq. (1) (whereby $H_u$ couples to up-type quarks and $H_d$
couples to down-type quarks and charged leptons) is not ``technically natural"
unless there is a symmetry $K$ under which $H_u$ and $\tilde{H}_d$
transform differently. The off-diagonal elements of $M^2_{ij}$ necessarily break
this symmetry. If $K$ is a global symmetry, then one can simply regard
the $d=2$ term $M_{ud}^2 H_u^{\dag} \tilde{H}_d$ as a soft breaking of $K$.
In that case, the pattern of Yukawa couplings in Eq. (1) is still technically natural.
It would be reasonable to assume that the natural scale of $M^2_{ud}$ was the same as
that of $M^2_{uu}$ and $M^2_{dd}$.

On the other hand, one might imagine that $K$ is spontaneously broken. For example,
suppose $M^2_{ud}$ arises from a $K$-invariant term $\epsilon H_u^{\dag} \tilde{H}_d S^2$,
where $S$ has a potential $V(S) = \frac{1}{4} \lambda_S (|S|^2)^2 + \mu^2_S |S|^2$.
Then it would be consistent with our approach to say that $\mu^2_S$ scanned and had
a natural scale $M^2_*$. In this scenario, there are two equally probable cases,
$\mu_S^2 >0$ and $\mu_S^2 < 0$. For $\mu^2_S <0$, the natural scale of $M_{ud}^2$ is
$\epsilon \lambda_S^{-1} M_*^2$. If $\epsilon \lambda_S^{-1}$ is small, the
probability distribution $P(x_u, x_d)$ is suppressed when $v_u$ and $v_d$ are comparable
(more precisely, when they are within a factor of $\epsilon \lambda_S^{-1}$ of each
other), i.e. near the line $x_u = \overline{x_d}$ in Fig. 3. This would not qualitatively
affect our analysis unless $\epsilon \lambda_S^{-1}$ were very small (less than about $10^{-2}$),
in which case the potentially viable small-$c$- quark and small-$s$-quark regions would
start to have probability densities rivaling that of our domain. If $\mu_S^2 >0$, then
$M^2_{ud} = 0$ and quite a different situation results. A single fine-tuning to
make the weak scale small would leave either $M_{uu}^2$ or $M_{dd}^2$ of order
$+ M_*^2$. If $M_{uu}^2 \sim + M^2_*$, the $u$, $c$, and $t$ quarks would have
negligible masses, while the down quark masses would be set by the scale of weak
interaction breaking. We have not investigated the viability of such domains.

In our discussions of probability distributions, we have not considered
the renormalization group running of $\lambda$ and $\mu^2$ from the scale
$M_*$ down to the scale $v$.  This would affect the probability distributions by
factors of order 1. However, it should not qualitatively affect our discussion.

\section{Simultaneous tuning of the Higgs mass matrix and $\Lambda$}

In section 2.4, two anthropic constraints on the relation between the
mass of the electron and the cosmological constant were discussed.
These eliminate as non-viable the great majority of the positive-$\mu^2$
parameter space.   However, in order to express these as anthropic
constraints on the Higgs mass parameters (or equivalently on
$x_u = \ln (m_u/M_*)$ and $x_d = \ln (m_d/M_*)$) one must do
an analysis that simultaneously takes into account the probability
distributions of $x_u$, $x_d$ and $\rho_{\Lambda}$.  That is done in
this section.  Both of the anthropic constraints on $\rho_{\Lambda}$
discussed in section 2.4 and given in Eqs. (4) and (5) require $\rho_{\Lambda}$
to be less than bounds that are proportional to $m_e^3$ and thus to $e^{3 \overline{x_d}}$.
If we assume that the probability distribution of $\rho_{\Lambda}$
among domains is flat in the interval $[-M_{P \ell}^4,
+M_{P \ell}^4]$, then the probability distribution given in Eq. (20)
gets modified by a changed normalization factor and by an extra factor of
$e^{3 \overline{x_d}}$ to account
for the tuning of the cosmological constant required to satisfy Eq. (4) or  (5):

\begin{equation}
P'(x_u, x_d) \; dx_u \; d x_d  \propto
\frac{ e^{2 x_u + 2 \overline{x_d}}}
{(e^{2 x_u} + e^{ 2 \overline{x_d}})^{5/2}} \; e^{3 \overline{x_d}} \; dx_u \; d\overline{x_d}.
\end{equation}

\noindent
The prime on this probability density indicates that this is a conditional
probability: it is the probability density for being at $(x_u, x_d)$ subject
to the condition that $\rho_{\Lambda}$ has been tuned to a small enough
value to satisfy either Eq. (4) or Eq. (5).  If the condition is that
Eq. (4) be satisfied, we will call this $P'_{planet}$, if the condition is
that Eq. (5) is satisfied, we will call it $P'_{structure}$.

The constant of proportionality in Eq. (21) depends on which
condition is being enforced. To determine the constants of
proportionality, it is convenient to start by finding the points in
the positive-$\mu^2$ and negative-$\mu^2$ regions of the $x_u-x_d$
plane where $x_u = x_d$ and where the Higgs mass matrix $M^2_{ij}$
is fine-tuned to the same degree as in our domain (without any condition
on $\rho_{\Lambda}$).  We will call
these point $S_+$ and $S_-$. (The $S$ stands for symmetric, since
$x_u = x_d$ there, and the subscript tells whether it lies in the
$\mu^2 >0$ or $\mu^2 <0$ region.)  These points are shown in Fig. 7,
where the point representing our domain is labeled D.
\begin{figure}[h]
\begin{center}
\includegraphics{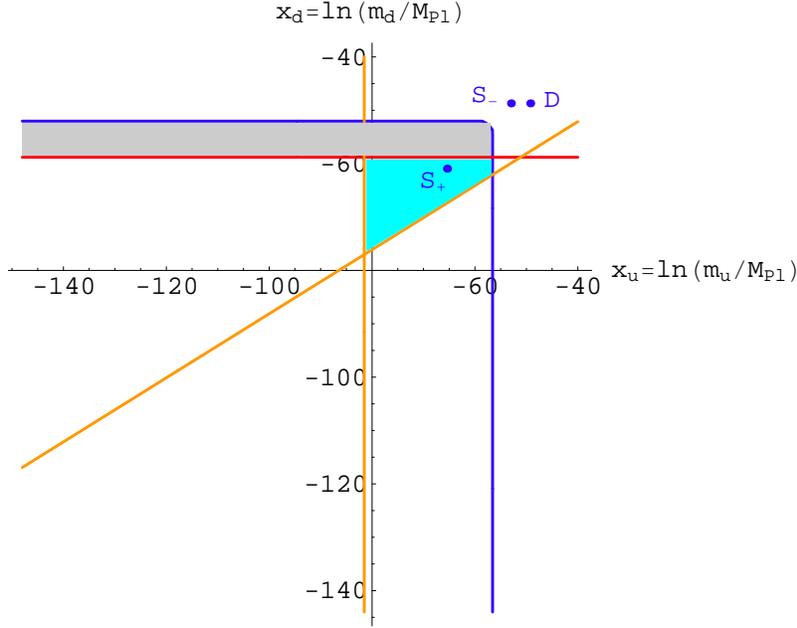}
\caption{The $\mu^2 >0$ region (approximately $v_u, v_d < f_{\pi}$) is 
below and to the left of the blue curve.
The arguments of section 2.4 rule out all but the shaded blue region.}
\label{Fig7}
\end{center}
\end{figure}
Suppose that
Eq. (5) implies that at $S_+$ the cosmological constant has to satisfy
$\rho_{\Lambda} (S_+) <  p_{structure} \; \rho_{\Lambda 0}$, where $\rho_{\Lambda 0}$
is the value observed in our domain, i.e. that $\rho_{\Lambda}$ must be
fine-tuned at $S_+$ to be less than it is in our domain by a
factor smaller than $p_{structure}$. On the other hand, the Higgs mass
parameters are equally fine-tuned at
$S_+$ and in our domain, by definition of $S_{\pm}$.  So the conditional probability
density $P'_{structure}$ at
$S_+$ divided by the conditional
probability density $P'$ in our domain (where $\rho_{\Lambda}$
is tuned to $\rho_{\Lambda0}$) is just given by $p_{structure}$.
Therefore, if we define
the shifted variables $\tilde{x_u} \equiv x_u - x_u(S_+)$ and
$\tilde{x_d} \equiv x_d - x_d(S_+)$, we may write

\begin{equation}
\frac{P'_{structure} (x_u, x_d)}{P'(D)} < p_{structure}
\; \frac{2^{5/2} e^{2 \tilde{x_u} +
5 \tilde{x_d}}}{(e^{2 \tilde{x_u}} + e^{2 \tilde{x_d}})^{5/2}}.
\end{equation}

\noindent
Note that at $S_+$, where $\tilde{x_u} = \tilde{x_d} = 0$,
the right-hand side of the
equation is just $p_{structure}$. In order to compute the coefficient $p_{structure}$,
it is necessary to determine the value of $m_e(S_+)$ and substitute it
into Eq. (5).

At the point $D$, i.e. in our domain, one has $v_u = v_{u0} \simeq
174$ GeV, $v_d = v_{d0} \simeq 4$ GeV (if we assume $Y_b \simeq Y_t
\simeq 1$), and $m_e = m_{e0} = Y_e v_{d0} \simeq Y_e$ (4 GeV).
Since $P(S_+) = P(S_-)$, Eq. (16) implies that $v_u(S_-)  = v_d(S_-)
\simeq \sqrt{2} v_{d0} \simeq 4 \sqrt{2}$ GeV, and therefore $v(S_-)
\simeq 8$ GeV; so that $m_e(S_-) = (Y_e/\sqrt{2}) v(S_-) \simeq Y_e
(4 \sqrt{2}$ GeV) $\simeq \sqrt{2} m_{e0} = 0.7$ MeV.  On the other
hand, at the point $S_+$, the electron mass is given by $m_e(S_+) =
(Y_e/\sqrt{2}) \; (f_{\pi}^3/\mu^2(S_+))$.
But by definition, $\mu^2(S_+) = \mu^2(S_-)$ and
therefore $\sqrt{\mu^2(S_+)/\lambda} = \sqrt{\mu^2(S_-)/\lambda} =
v(S_-) \simeq 8$ GeV.  Taking $\lambda \simeq 1$ we have $m_e(S_+) \simeq
m_e(S_-) (f_{\pi}/(8 {\rm GeV}))^3 \simeq 1$ eV.
Substituting this into Eq. (5), one has
$\rho_{\Lambda}(S_+) < 10^{-7} \rho_{\Lambda 0}$.
In other words,

\begin{equation}
p_{structure} \cong 10^{-7}.
\end{equation}

The constraint on $\rho_{\Lambda}$ coming from Eq. (5) only applies if
$m_e > 20$ eV. The region with $m_e > 20$ eV lies above the horizontal
straight red line in Fig. 7. The region of positive $\mu^2$ corresponds
to the area below and to the left of the blue curve in Fig. 7.
(This curve arises from the fact that for very small
positive $\mu^2$, the value of
$\langle H_< \rangle$ is set by the competition between
the quartic term $(\lambda/4) H_<^4$ and the linear term
for $H_<$ coming from the QCD condensates of $\overline{q}q$.
This gives a maximum value for
$\sqrt{v_u^2 + v_d^2} = \langle H_< \rangle \simeq f_{\pi}$
in the positive-$\mu^2$ region.)  It is easy to see from
Eqs. (22) and (23) that the entire region satisfying both
$\mu^2 >0$ and $m_e > 20$ eV (shaded gray in Fig. 7) is much more fine-tuned than our
domain, i.e.
$P'_{structure}/P'(D)  \ll 1$.

Turning now to the condition that planets be able to
form (Eq. (4)), the calculation is completely parallel to the one
just done.
To find $p_{planet}$ one substitutes $m_e(S_+)$ into
Eq. (4) and divides by $\rho_{\Lambda0}$, which gives

\begin{equation}
\frac{P'_{planet} (x_u, x_d)}{P'(D)} < p_{planet}
\; \frac{2^{5/2} e^{2 \tilde{x_u} +
5 \tilde{x_d}}}{(e^{2 \tilde{x_u}} + e^{2 \tilde{x_d}})^{5/2}}.
\end{equation}

\noindent
and

\begin{equation}
p_{planet} \simeq 3 \times 10^{13} \; A,
\end{equation}

\noindent
where $A$ is the average atomic weight of a nucleus in the planet.
It is easy to see that the region to the left of the vertical
orange line or below the slanted orange line are more fine-tuned
than our domain, i.e. $P'_{planet}/P'(D) <1$.  The shape of the
orange line arises from the fact that when $\tilde{x_d} > \tilde{x_u}$,
the expression in Eq. (24) is approximately proportional to simply
$e^{2 \tilde{x_u}}$, whereas when $\tilde{x_u} >
\tilde{x_d}$, the expression in Eq. (24) is approximately
proportional to
$e^{- 3 \tilde{x_u} + 5 \tilde{x_d}}$.

The two arguments given in section 2.4 have excluded (as
being more fine-tuned than our domain) the whole of
the positive-$\mu^2$ region except for the region shaded
light blue in Fig. 7.  However, that is not
good enough. It is necessary to exclude the {\it entire}
positive-$\mu^2$ region either as not viable or as more
improbable than our domain, if the anthropic explanation of
the observed values of $m_u$ and $m_d$ is to be tenable.
Other arguments that were suggested in \cite{abds} hold
some promise of being able to do this.  For example, as
noted there, it might be that the peculiarities of nuclear
physics in the positive-$\mu^2$ region may lead to runaway
nucleosynthesis.  However, the situation for $\mu^2 >0$ has
many unclear aspects, and more thought must be given to it.

\section{Conclusions}

We have investigated the possibility that the entire mass matrix of
the Higgs fields in a two-Higgs-doublet model varies among domains
of a ``multiverse".  This means that both the weak scale
$v \equiv \sqrt{v_u^2 + v_d^2}$ and $\tan \beta \equiv v_u/v_d$
vary among domains.  Making plausible assumptions about probability
distributions of parameters in the ``landscape" and about the requirements
for life, we have found that almost all of the $\ln m_u - \ln m_d$ space
is either anthropically non-viable or much more fine-tuned
(and therefore more rare in the multiverse) than our domain.  The
potentially viable regions are islands in parameter space.
One of these islands is ours, and it is a very small island indeed,
as seen in Figs. 3 and 4.
If this is the only island of viability, there is the possibility of accounting
for two otherwise not easily understood facts about the world we observe,
namely that the weak scale is comparable to the strong scale, and
that $m_u/m_d \sim 1$.

Whether this is in fact the only island of viability is not yet clear.
There are possible islands of viability in the ``light-$c$-quark" and
``light-$s$-quark" regions.  However, even if islands of viability do exist in
those parts of parameter space they are less probable in the landscape than
the island of viability in which our domain is located, according to the
analysis of section 3. More significant is the possibility of an island
of viability in the $\mu^2 >0$ region of parameter space.

An interesting new twist has been given to the discussion of
the positive-$\mu^2$ case by the discovery of dark matter/cosmological
constant.  This discovery ties together the anthropic analysis of the
Higgs mass given in \cite{abds} and \cite{hogan} and the anthropic
analysis of the cosmological constant given in \cite{weinberg}.
We have seen that some of the $\mu^2 >0$ constraints are invalidated
by the existence of dark energy, but other and equally strong
constraints that depend on the existence of dark energy come into play.
However, the constraints we have discussed are not by themselves enough to
rule out the entire $\mu^2 >0$ region as non-viable. This seems to be the main
loophole at present in the attempt to explain $m_u \sim m_d$ and the value of
the weak scale anthropically. However, it seems probable that much stronger
arguments exist for the case of
$\mu^2 >0$, and that the entire positive-$\mu^2$ region of parameter space is
non-viable.

We have assumed that only the dimensionful parameters of the
Lagrangian ``scan", i.e. vary in the multiverse, and that only they are
anthropically tuned.  This seems like a comparatively
conservative assumption, since the dimensionful parameters
of our current theories are the least well understood, are
the most (apparently) fine-tuned, and are the most amenable
to anthropic explanation. However, obviously, other assumptions
are possible.  The question for us is whether there exists at
least some set of plausible assumptions  under which the observed
relation $m_u \sim m_d$ can be anthropically explained.

\section*{Acknowledgements}

We acknowledge useful conversations with D. Seckel, Ilia Gogoladze, and Tian-jun Li.
This research was supported by the DOE grant DE-FG02-91ER40626.

\end{document}